\documentclass{article}
\usepackage{ijcai25}

\usepackage{times}
\usepackage{soul}
\usepackage{url}
\usepackage[hidelinks]{hyperref}
\usepackage[utf8]{inputenc}
\usepackage[small]{caption}
\usepackage{graphicx}
\usepackage{amsmath}
\usepackage{amsthm}
\usepackage{booktabs}
\usepackage{algorithm}
\usepackage{algorithmic}
\usepackage[switch]{lineno}
\usepackage{multirow}
\usepackage{subfigure}
\usepackage{bbding}
\usepackage{makecell}
\usepackage{hyperref}
\usepackage{amsfonts}
\usepackage{amssymb}
\usepackage{tikz} 
\usepackage{xcolor} 

\title{ListenNet: A Lightweight Spatio-Temporal Enhancement Nested Network for Auditory Attention Detection}

\author{
    Author Name
    \affiliations
    Affiliation
    \emails
    email@example.com
}

\author{
Cunhang Fan\and
Xiaoke Yang\and
Hongyu Zhang\and
Ying Chen\and
Lu Li\and
Jian Zhou\And
Zhao Lv\thanks{Corresponding Author}\\
\affiliations
Anhui Province Key Laboratory of Multimodal Cognitive Computation, School of Computer Science and Technology, Anhui University\\
\emails
\{cunhang.fan, jzhou, kjlz\}@ahu.edu.cn,
\\\{e22201014, e22201103, e23201035, e12314059\}@stu.ahu.edu.cn}

\begin{document}

\maketitle


\begin{abstract}
    Auditory attention detection (AAD) aims to identify the direction of the attended speaker in multi-speaker environments from brain signals, such as Electroencephalography (EEG) signals. However, existing EEG-based AAD methods overlook the spatio-temporal dependencies of EEG signals, limiting their decoding and generalization abilities. To address these issues, this paper proposes a Lightweight Spatio-Temporal Enhancement Nested Network (ListenNet) for AAD. The ListenNet has three key components: Spatio-temporal Dependency Encoder (STDE), Multi-scale Temporal Enhancement (MSTE), and Cross-Nested Attention (CNA). The STDE reconstructs dependencies between consecutive time windows across channels, improving the robustness of dynamic pattern extraction. The MSTE captures temporal features at multiple scales to represent both fine-grained and long-range temporal patterns. In addition, the CNA integrates hierarchical features more effectively through novel dynamic attention mechanisms to capture deep spatio-temporal correlations. Experimental results on three public datasets demonstrate the superiority of ListenNet over state-of-the-art methods in both subject-dependent and challenging subject-independent settings, while reducing the trainable parameter count by approximately 7 times. Code is available at: \href{https://github.com/fchest/ListenNet}{\textcolor{blue}{https://github.com/fchest/ListenNet}}.
\end{abstract}

\section{Introduction}
In multi-speaker environments, humans with normal hearing have the ability to focus on a specific speaker while ignoring interference from other sound sources, the phenomenon known as the cocktail party effect
\cite{cherry1953some}. The mechanism behind it is commonly referred to as selective auditory attention. This inherent ability plays a crucial role in human communication and has attracted growing interest in auditory attention detection (AAD), which aims to localize the attended speaker using brain signals \cite{dai2018neural}. AAD could potentially enhance the design of human-centered intelligent interaction systems, such as hearing aids. 

Neuroscientific studies have demonstrated a nonlinear relationship between auditory attention and brain activity \cite{choi2013quantifying,mesgarani2012selective}, which involves higher cognitive processing in the cerebral cortex. Electroencephalography (EEG) signals
are widely used due to their non-invasive nature, ease of acquisition, and high temporal resolution \cite{de2020machine,fan2024light}. Spatio-temporal patterns of EEG reveal attentional regulation during selective listening \cite{tune2021neural}. Findings of the inter-subject correlation (ISC) suggest that EEG signals are synchronized across subjects during perception of the same naturalistic visual and narrative speech stimuli \cite{dmochowski2014audience,shen2022contrastive}. Taking this perspective, EEG signals exhibit temporal correlations, spatial correlations across channels, and spatio-temporal dependencies, which could provide valuable information for discriminating different attention states and advancing robust AAD methods. 

\begin{figure}[tb]
\centering
\includegraphics[width=1\columnwidth]{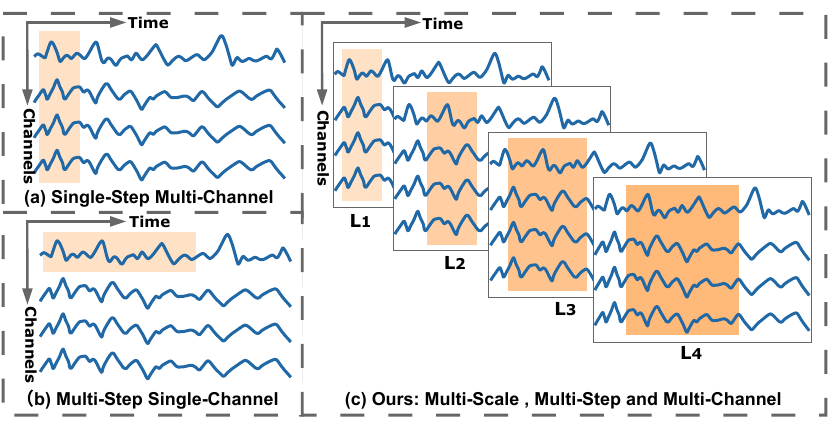}
\caption{Spatio-temporal modeling methods for AAD. Existing methods typically treat space and time separately, processing them from (a) to (b). The proposed ListenNet introduces a multi-scale of temporal patterns, as shown in (c), by considering cross-channel dependencies, temporal dynamics, and spatio-temporal dependencies for a more comprehensive modeling approach.}
\label{fig:1}
\end{figure}

Despite significant progress made by existing EEG-based AAD methods, three major challenges still limit their performance and practical application. Firstly, many existing methods have made substantial strides in spatio-temporal modeling, effectively capturing dynamic spatial patterns, leading to improved detection performance. These methods typically treat space and time separately, as shown in Figure \ref{fig:1}~(a) and (b). Spatial dependencies are captured independently, and temporal dependencies are subsequently extracted. However, these methods overlook the temporal context under dynamic time conditions, as well as the spatio-temporal dependencies across different channels during auditory stimulus processing. Secondly, the individual differences and the non-stationary characteristics of EEG signals lead to significant performance degradation when applying AAD methods across subjects. \cite{cai2024robust,fan2024dgsd} effectively leverage individual-specific features to  demonstrate strong performance in the subject-dependent setting, but they lack good generalization ability, which makes it difficult to develop subject-independent robust methods. Lastly, the pursuit of accuracy in current methods \cite{jiang2022detecting,ni2024dbpnet} leads to large model sizes and high computational complexity, which are often attributed to complex feature extraction methods and transformer attention mechanisms, making them impractical for low-power devices. 

To address these issues, this paper proposes a \textbf{Li}ghtweight \textbf{S}patio-\textbf{T}emporal \textbf{E}nhancement \textbf{N}ested \textbf{Net}work (ListenNet) with low parameter count and computational complexity. As shown in Figure \ref{fig:1}~(c), it captures multi-channel spatio-temporal dependencies 
 and multi-scale dynamic temporal patterns, ensuring high accuracy and strong generalization. Specifically, the proposed ListenNet consists of three components: (1) \emph{Spatio-temporal Dependency Encoder (STDE)} captures consecutive time steps and multi-channel features, differing from previous studies that first focus on channel features. It expands the input EEG signals within each channel to capture temporal dependencies and extracts spatial features both within and across channels, enhancing spatio-temporal representation capacity. (2) \emph{Multi-scale Temporal Enhancement (MSTE)} captures temporal dependencies at multiple time scales, adding dynamic temporal context to build robust temporal embeddings. (3) \emph{Cross-Nested Attention (CNA)} groups spatio-temporal features in parallel, extracts sub-feature context, and recalibrates weights by encoding global information, enhancing deep spatio-temporal correlations. Finally, the effective features are passed to a classifier to predict the subject's attended speaker. The major contributions of this work are summarized as follows: 

\begin{itemize}
  \item[$\bullet$] The proposed ListenNet overcomes the performance and efficiency limitations of existing methods for AAD by efficiently capturing spatio-temporal dependencies in both subject-dependent and subject-independent settings. 
  \item[$\bullet$] A novel MSTE module is designed to efficiently extract multi-channel dependencies across multiple scales and time steps to integrate multi-level features, enhancing and complementing robust temporal representations.
  \item[$\bullet$] Experimental results show that ListenNet achieves outstanding accuracy while reducing the trainable parameter count by approximately 7 times. Specifically, it surpasses the best baseline by 6.1\% on the DTU dataset under the subject-dependent setting and by 8.2\% on the KUL dataset under the subject-independent setting, all within a 1-second decision window. 
  
\end{itemize}

\section{Related Works}

For spatial dependency modeling, existing methods are divided
into physical and dynamic dependencies. \cite{cai2021low,jiang2022detecting} project differential entropy (DE) features in the frequency domain onto 2D topological maps using the known electrode positions to calculate spatial dependency based on physical distance and achieve good performance. Although physical dependency conforms to prior
physiological paradigms, the electrode positions relations between channels cannot be directly equated to their functional connections \cite{liu2024vsgt}. Currently, some researchers autonomously learn spatial dependency relationships during training. \cite{fan2024dgsd} extracts DE features as nodes to construct graph neural networks (GNN) and utilize an updated parameter matrix to represent spatial dependency. \cite{su2021auditory,cai2023brain,cai2024robust} design channel-wise attention mechanisms that learn to assign distinct weights to capture spatial patterns. \cite{ni2024dbpnet} utilizes a dual-branch approach to extract features from the temporal and frequency domains in parallel. For the frequency branch, it projects DE onto 2D maps and uses their topological patterns. For the temporal branch, the transformer encoder embeds a single cross-channel time step as an input token to autonomously learn features. The current state-of-the-art (SOTA) study \cite{yan2024darnet} employs spatial convolution operations across all channels to effectively capture spatial dependencies, resulting in competitive AAD performance. 

For temporal dependency modeling, existing methods typically capture temporal dependencies using convolutional neural networks (CNN) and attention mechanisms. \cite{monesi2020lstm} independently uses long short-term memory (LSTM) networks to capture dependencies within EEG signals and achieve decent decoding performance. \cite{vandecappelle2021eeg} applies a simple one layer CNN model to directly process EEG data, where the time series are reduced to a single value. \cite{su2022stanet} sequentially processes temporal information after spatial attention, multiplying attention maps with EEG signals for adaptive feature refinement. \cite{wang2023eeg} utilizes a temporal attention mechanism after GNN that assigns varying weights to a sequence of EEG signals, enabling the capture of the complex temporal dynamics and enhancing the detection of even subtle changes in attentional states over time. Recently, \cite{eskandarinasab2024gru} employs gated recurrent units (GRU) and CNN to consider both historical and new temporal information when calculating the current state value, thereby inferring the temporal dependencies between time steps.

The methods mentioned above often focus separately on spatial and temporal features, or adopt a two-step processing strategy in which spatial dependencies are captured, followed by the modeling of temporal dependencies. However, these approaches tend to overlook the rich temporal contextual information under dynamic time conditions, as well as the spatio-temporal distribution characteristics of different brain regions during the reception, processing, and response to auditory stimuli. As a result, the failure to capture critical spatio-temporal dependencies significantly limits model performance.

\begin{figure*}
\centering
\includegraphics[width=0.9\textwidth]{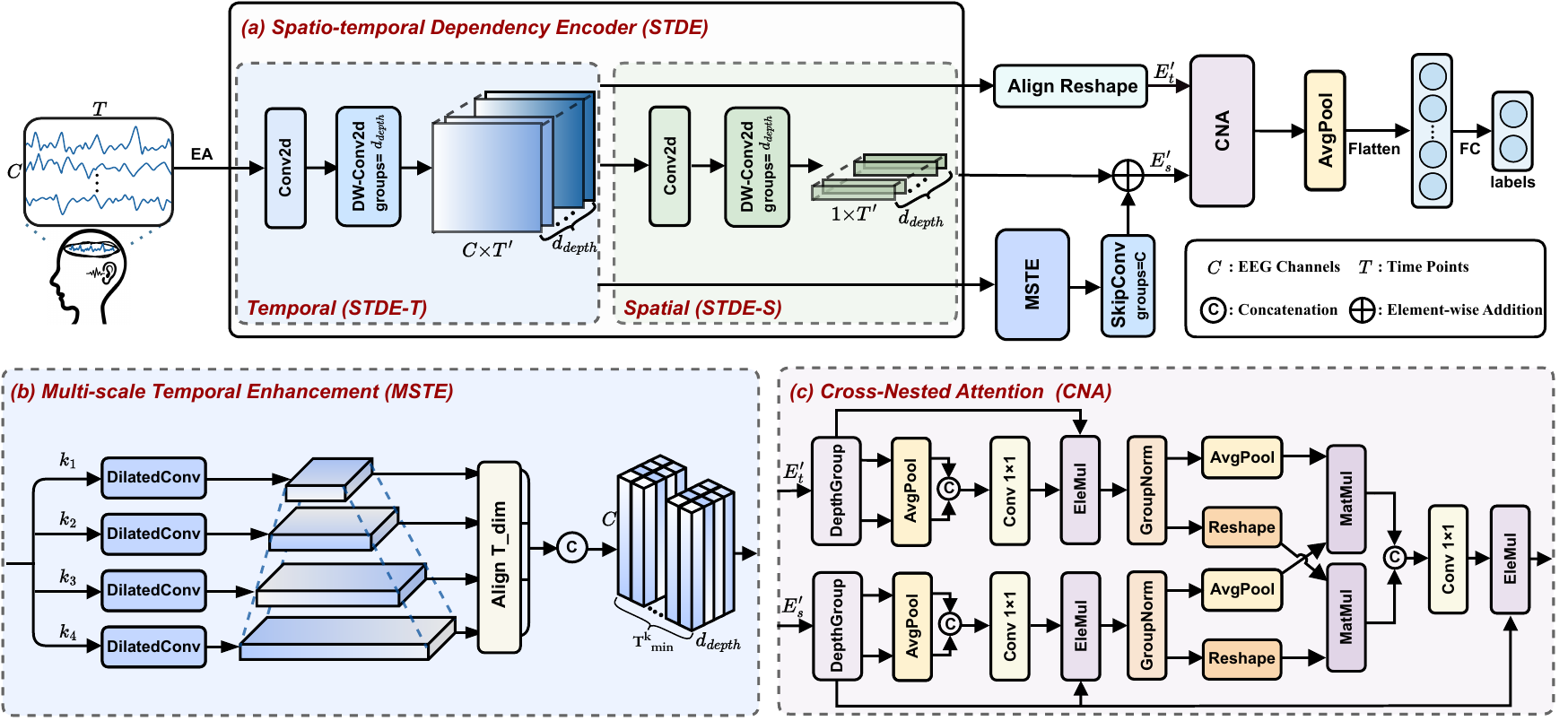}
\caption{The overall structure of our ListenNet for AAD consists of three modules: (a) STDE module, (b) MSTE module, where \( k_i \) (\(i \in \{1, 2, 3, 4\}\)) represents the kernel size used in the dilated convolution, and (c) CNA module, where \(E_t'\) and \(E_s'\) are depth-aligned input feature maps.
The model inputs are normalized and Euclidean-aligned EEG signals, and the outputs are two predicted labels related to auditory attention obtained through a classifier applied to the CNA output features.}
\label{fig:2}       
\end{figure*}

\section{The Proposed ListenNet Method}
The proposed ListenNet is designed to comprehensively integrate
spatio-temporal dependencies in EEG signals, addressing the limitations of existing methods by modeling dependencies across both multiple channels and time scales. Figure \ref{fig:2} illustrates the overall structure of ListenNet. The method will be specified in the following subsections.

Given the EEG data split by a moving window, a series of decision windows is obtained, each containing a short time segment of EEG signals. Consider the original EEG data of a decision window represented by \( X = [x_1,...,x_i,...,x_T] \in \mathbb{R}^{C \times T} \), where \( C \) is the number of EEG channels and \( T \) is the length of the decision window. Here, \( x_i \in \mathbb{R}^{C \times 1} \) is the EEG data at the \( i \)-th window of \( X \). We aim to learn a representation \( F(\cdot) \), which maps \( x \) to the corresponding label \( y=F(x) \). Here, \( y \) denotes the locus (i.e., left or right) of auditory attention. Before inputting the EEG data into ListenNet, a Euclidean alignment (EA) method \cite{miao2022priming} is employed, which standardizes the EEG data by calculating the average covariance matrix to extract shared features from the data across different brain states.
\(\tilde{X} \in \mathbb{R}^{C \times T}\) is obtained by normalizing and aligning \(X\).

\subsection{Spatio-temporal Dependency Encoder (STDE)} 

EEG signals are derived from different brain regions and exhibit dynamic changes in connectivity patterns between brain regions over time. Previous studies neglect the spatio-temporal characteristics of EEG signals. Meanwhile, as networks become increasingly complex \cite{zhang2023learnable,chen2023auditory,niu2024auditory}, the limited size of EEG data makes these networks prone to overfitting. CNN-based networks have demonstrated sufficient feature extraction capabilities in brain-computer interface (BCI) tasks \cite{lawhern2018eegnet,miao2023lmda}. Considering these characteristics, we design a spatio-temporal dependency encoder to extract robust dynamic patterns using depthwise separable convolutions, which consists of the temporal feature component (STDE-T) and the spatial feature component (STDE-S), as shown in Figure \ref{fig:2}~(a).

Firstly, STDE-T extracts dynamic features from EEG signals through temporal convolution layers, capturing temporal dependencies and constructing the temporal patterns \(E_t\). This can be expressed as:
\begin{equation}
E_t = GELU(DepthwiseConv(Conv(\tilde{X})))
\end{equation}
where \(E_t\in \mathbb{R}^{d_{\text{depth}} \times C \times T'}\), $Conv(\cdot)$ represents convolutional filters with a \( 1 \times 1 \) kernel size to perform spatio-temporal reshaping on the input signals. $DepthwiseConv(\cdot)$ performs convolution independently on each input channel along the time dimension with a kernel size \( 1 \times k_{\text{0}} \) and a group size \(d_{\text{depth}}\), followed by the $GELU(\cdot)$ activation function.

Subsequently, STDE-S encodes the spatial distribution information across all channels through spatial convolution layers, capturing the spatial distribution features \(E_s\) from EEG signals, which facilitate a comprehensive understanding of the brain's activity patterns in response to various auditory stimuli. This can be expressed as:
\begin{equation}
E_s = GELU(DepthwiseConv(Conv({E_t})))
\end{equation}
where \(E_s \in \mathbb{R}^{d_{\text{depth}} \times 1 \times T'} \), $Conv(\cdot)$ represents convolutional filters with a \( 1 \times 1 \) kernel size for initial channel mapping and achieving channel-wise feature fusion. $DepthwiseConv(\cdot)$ performs convolution to capture inter-channel dependencies with a \( C \times 1 \) and a group size \(d_{\text{depth}}\), with the $GELU(\cdot)$ activation function. We integrate the spatial distribution features with the temporal patterns to form a comprehensive spatio-temporal embedding \( E_s \).

\subsection{Multi-scale Temporal Enhancement (MSTE)}
The auditory system is sensitive to the temporal patterns \cite{puffay2022relating}. Inspired by the concept of multi-scale modeling \cite{wu2020connecting,fan2024msfnet}, we propose a novel MSTE module. As shown in Figure \ref{fig:2}~(b), the module captures dynamic brain activity across multiple time scales, offering a comprehensive representation of temporal patterns.

MSTE integrates dilated convolutions with the inception strategy to capture temporal features across multiple scales, thereby enabling a more comprehensive representation of multi-level temporal dependencies and enhancing the modeling of complex temporal patterns. The dilated convolution filters use four different kernel sizes to capture patterns at different time scales, with same dilation factor progressively expanding the effective receptive field. This enables the module to more efficiently capture both fine-grained and long-term temporal dependencies without increasing the number of parameters. Formally, the Inception strategy is combined with dilated convolutions to capture multi-scale temporal features. Given the input from the temporal convolution layers, the module applies four convolutional filters, each with a fixed dilation factor, to extract multi-scale temporal features. The outputs are truncated to match the size of the largest kernel, concatenated along the channel dimension, and normalized using batch normalization, ultimately generating the multi-scale feature map. The above process can be formulated as:
\begin{equation}
U = [\text{DilatedConv}_{1 \times k}(E_t) \mid k \in \{k_1, k_2, k_3, k_4\}]
\label{eq:dilated_inception}
\end{equation}
where $U \in \mathbb{R}^{d_{\text{depth}} \times C \times T^{k}_{\min}}$, and $T^{k}_{\min}$ represents the minimum time dimension among the outputs. 
$[\cdot]$ represents concatenation operation, and $\textit{DilatedConv}_{1 \times k}(\cdot)$ is implemented as a set of dilated convolutions with $k \in \{k_1, k_2, k_3, k_4\}$. 
For each kernel size $k$, a convolution is applied along the temporal dimension with a fixed dilation factor $d$.

The skip connection is implemented using a depthwise convolution with a kernel size of \(C\times 1 \) and a group size \(d_{\text{depth}}\). These transform spatial information while preserving channel structure and standardize the sequence length for consistent transmission to the output module. The features are resized via bilinear interpolation to match the dimensions required by the subsequent layer, resulting in \( S \in\mathbb{R}^{d_{\text{depth}} \times 1 \times T'} \), which is then added to \( E_s \), producing a robust representation of spatio-temporal dynamics \( E_s' \in \mathbb{R}^{d_{\text{depth}} \times 1 \times T'} \).


\subsection{Cross-Nested Attention (CNA)}
The multi-head attention mechanism in transformer models achieves significant results but incurs high computational cost. Inspired by the parallel strategy for cross-dimensional spatial information aggregation \cite{wang2020eca,ouyang2023efficient}, we propose a novel cross-nested attention module that efficiently integrates hierarchical spatio-temporal features and reduces computational cost.

CNA employs dual-branch decomposition and interactive enhancement, extracting deep spatio-temporal features through attention weighting. Prior to processing, the input temporal feature \(E_t\) is depth-aligned with \(E_s'\) to produce \(E_t' \in \mathbb{R}^{d_{\text{depth}} \times d_{\text{depth}} \times T'}\). As shown in Figure~\ref{fig:2}(c), both \(E_t'\) and \(E_s'\) are divided into \(G\) groups along the depth dimension, where \(G = \left\lfloor d_{\text{depth}} / 2 \right\rfloor\), and \(\left\lfloor \cdot \right\rfloor\) denotes the floor operation. The dimension-adjusted features are denoted as \(F_t\) and \(F_s\), respectively. Then, a dual-branch spatio-temporal module is applied to decompose and capture global information in both directions, producing two enhanced features, \(F_1\) and \(F_2\), as formulated below:
\begin{equation}
    \begin{aligned}
        F_1 &= GN\left( F_t \odot \sigma\left( AAP_S(F_t) \right) \odot \sigma\left( AAP_T(F_t) \right) \right) \\
        F_2 &= GN\left( F_s \odot \sigma\left( AAP_S(F_s) \right) \odot \sigma\left( AAP_T(F_s) \right) \right)
    \end{aligned}
\end{equation}
where, \( GN(\cdot) \) denotes the group normalization operation, \( AAP_S(\cdot) \) denotes the spatial adaptive average pooling operation, \( AAP_T(\cdot) \) denotes the temporal adaptive average pooling operation, \( \sigma(\cdot) \) denotes the sigmoid activation function, and \( \odot \) denotes the element-wise multiplication operation.

To capture long-range dependencies and global context, global average pooling and softmax are applied to each input branch to produce attention vectors. These are reshaped and used to compute cross-attention maps with features from the opposite branch via matrix multiplication. The resulting maps are concatenated and passed through a shared \(1 \times 1\) convolution for feature fusion and dimensionality reduction, yielding the final attention weights \(W \in \mathbb{R}^{(B \times G) \times 1 \times 1 \times T'}\), with \(B\) denoting the batch size. Finally, the output deep spatio-temporal features \( E \in \mathbb{R}^{d_{\text{depth}} \times 1 \times T'} \) are obtained by applying element-wise multiplication between \(F_s\) and the sigmoid-activated \(W\).

\subsection{Classifier}
The classifier is designed to provide the final auditory attention results. Global average pooling is applied to reduce the dimensions of the features output by the CNA module. Then, the normalized feature maps are flattened into a 1D vector and fed into a fully connected layer to produce the final result. In the training stage, we apply the binary cross-entropy function to update the parameters.

\begin{equation}
    \mathcal{L} = -\frac{1}{N}\sum_{i=1}^{N}[y_i \cdot \log Q_i + (1 - y_i) \cdot \log (1 - Q_i)]
\end{equation}
where $y_i$ means the ground-truth label of $i$-th decision window, $N$ means the number of samples, and $Q_i$ is the corresponding possibility of predicted direction label with softmax function processing.

\section{Experiment}
\subsection{Datasets}
We evaluate ListenNet on three publicly available datasets: KUL \cite{das2019auditory,das2016effect}, DTU \cite{fuglsang2018eeg,fuglsang2017noise}, and AVED \cite{zhang2024based}. We summarize the details of the above datasets in Table \ref{tab:1}.

1) \textbf{KUL:} This dataset consists of 16 normal-hearing subjects, with 64-channel EEG data recorded. Each subject was instructed to attend to one of two competing voices from either the 90$^{\circ}$ left or right. Each subject completed 8 trials, each lasting 6 minutes.


2) \textbf{DTU:} This dataset consists of 18 normal-hearing subjects, with 64-channel EEG data recorded. Each subject was instructed to perform a target speaker tracking task in an environment with reverberation and dynamic background noise interference, attending to one of two competing voices from speakers positioned at a 60$^{\circ}$ relative to the subject. Each subject completed 60 trials, each lasting 50 seconds.


3) \textbf{AVED:} This dataset consists of 20 normal-hearing subjects, with 32-channel EEG data recorded. Subjects were evenly divided into two experimental conditions: audio-only and audio-visual, with 10 subjects in each condition. Each subject was instructed to attend to one of two competing voices from either the 90$^{\circ}$ left or right. In the audio-visual condition, subjects not only listened to the stories but also watched the video of the narrator they were instructed to focus on. Each subject completed 16 trials, each lasting 152 seconds.


\begin{table}[bt]
\centering
\resizebox{\columnwidth}{!}{
\begin{tabular}{cccccc}
\toprule[1pt]
\multirow{2}{*}{\textbf{Dataset}} & \multirow{2}{*}{\textbf{Scene}} & \multirow{2}{*}{\textbf{Subjects}} & \multirow{2}{*}{\textbf{Channels}} & \textbf{Stimulus} &\textbf{Duration} \\ &  &  &  &
\textbf{Direction}&\textbf{(minutes)} \\ \midrule
\vspace{0.5mm}  
KUL   &audio-only        & 16                & 64    & ±90$^\circ$                       & 48                      \\
\vspace{0.8mm}  
DTU    & audio-only         & 18                & 64   & ±60$^\circ$                       & 50                     \\
\vspace{0.5mm}  
\multirow{2}{*}{AVED}   & audio-only       &10                &32    & ±90$^\circ$                        &40   \\ & audio-visual &10 &32 &±90$^\circ$  &40 \\ \bottomrule[1pt]
\end{tabular}
}
\caption{Details of three EEG datasets used in experiments.}
\label{tab:1}
\end{table}
\setlength{\textfloatsep}{8pt} 

\subsection{Data Processing}
To eliminate artifact noise and obtain cleaner EEG signals, specific preprocessing steps are applied to the three datasets to ensure consistency and comparability across experiments. For the KUL dataset, EEG signals are band-pass filtered (0.1--50 Hz) to remove irrelevant frequencies and downsampled to 128 Hz. For the DTU dataset, 50 Hz line noise and power line interference are filtered out, followed by downsampling to 128 Hz and high-pass filtering at 0.1 Hz. Eye artifacts are removed using joint decorrelation, and data are re-referenced to the average EEG channel response. For the AVED dataset, 50 Hz power line interference is removed, and the signals are band-pass filtered (0.1--50 Hz) and downsampled to 128 Hz. Subsequently, ocular and muscle artifacts are eliminated using independent component analysis (ICA). Finally, all EEG channels were re-referenced.


To evaluate ListenNet, we compare it with other SOTA AAD methods under both subject-dependent and the more challenging subject-independent settings. Specifically, four open-source models are selected as baselines: SSF-CNN \cite{cai2021low}, MBSSFCC \cite{jiang2022detecting}, DBPNet \cite{ni2024dbpnet}, and DARNet \cite{yan2024darnet}.

\subsection{Implementation Details}

We evaluate the performance of ListenNet on KUL, DTU, and AVED datasets under both subject-dependent and subject-independent settings. For the subject-dependent condition, each subject's data is split into training, validation, and test sets in an 8:1:1 ratio. The batch size is set to 32, the maximum number of epochs to 100, and an early stopping strategy is employed. Moreover, the model is trained using an Adam optimizer with a learning rate of 5e-4 and weight decay of 3e-4. For the subject-independent condition, the leave-one-subject-out (LOSO) cross-validation strategy is used. Namely, one subject’s EEG data constituted the testing data, and the remaining subjects’ EEG data constituted the training data. Here, the batch size is set to 128, with a maximum of 100 epochs. An Adam optimizer is also used with a learning rate of 1e-3 and a weight decay of 3e-4.

The following describes the implementation details, including the training settings and network configuration. The hyperparameters of ListenNet are consistently fixed across the three datasets to ensure a fair comparison of its generalizability. For STDE, the kernel size \(k_{\text{0}}\) is set to 8, and the group size \(d_{\text{depth}}\) is set to 16. For MSTE, the kernel sizes used in the 2D dilated convolutional filter are \(k \in \{1, 2, 3, 5\}\), and the dilation factor \(d\) is set to 1. Consequently, the number of groups \(G\) in CNA is configured as 8. All experiments are conducted using PyTorch on an RTX 4090 GPU.

\section{Results}

\subsection{Comparison with Prior Art}
\begin{table*}[tb]
\centering
\resizebox{0.9\textwidth}{!}{%
\begin{tabular}{cccccccc}
\toprule[1pt]
\multirow{2}{*}{\textbf{Dataset}} & \multirow{2}{*}{\textbf{Scene}} & \multirow{2}{*}{\textbf{Model}}                         & \multicolumn{3}{c}{\textbf{Subject-Dependent}} & \multicolumn{2}{c}{\textbf{Subject-Independent}} \\ \cmidrule{4-8} 
\multicolumn{2}{c}{}                                                                                                                  &                                                & \multicolumn{1}{c}{0.1-second} & \multicolumn{1}{c}{1-second} & \multicolumn{1}{c}{2-second}       & \multicolumn{1}{c}{\centering 1-second}  & \multicolumn{1}{c}{\centering 2-second} \\ \midrule
\multirow{8}{*}{KUL} & \multirow{8}{*}{audio-only} & CNN \cite{vandecappelle2021eeg} & 74.3 & 84.1 & 85.7 & 56.8 ± 5.58  & 59.5 ± 8.21 \\
\multicolumn{2}{c}{} & SSF-CNN \cite{cai2021low} & 76.3 ± 8.47 & 84.4 ± 8.67 & 87.8 ± 7.87 & 59.3 ± 6.69  & 60.8 ± 8.40 \\
\multicolumn{2}{c}{} & MBSSFCC \cite{jiang2022detecting} & 79.0 ± 7.34 & 86.5 ± 7.16 & 89.5 ± 6.74 & 62.7 ± 8.08  & 64.7 ± 8.62 \\
\multicolumn{2}{c}{} & EEGraph \cite{cai2023brain} & 88.7 ± 6.59 & 96.1 ± 3.22 & 96.5 ± 3.34 & \multicolumn{1}{c}{-}  & \multicolumn{1}{c}{-} \\
\multicolumn{2}{c}{} & DGSD \cite{fan2024dgsd} & \multicolumn{1}{c}{-} & 90.3 ± 7.29 & 93.3 ± 6.53  & 63.6 ± 8.00  & \multicolumn{1}{c}{-}\\
\multicolumn{2}{c}{} & DBPNet \cite{ni2024dbpnet} & 85.3 ± 6.22 & 94.4 ± 4.62 & 95.3 ± 4.63 & 61.1 ± 8.26  & 62.3 ± 7.37 \\
\multicolumn{2}{c}{} & DARNet \cite{yan2024darnet} & 89.2 ± 5.50 & 94.8 ± 4.53 & 95.5 ± 4.89 & 69.9 ± 11.82  & 71.9 ± 13.01 \\
\multicolumn{2}{c}{} & \textbf{ListenNet (ours)} & \textbf{92.5 ± 5.24} & \textbf{96.9 ± 3.01} & \textbf{97.3 ± 2.62} & \textbf{78.1 ± 13.50}  & \textbf{79.6 ± 14.60} \\

\midrule
\multirow{8}{*}{DTU} & \multirow{8}{*}{audio-only} & CNN \cite{vandecappelle2021eeg} & 56.7 & 63.3 & 65.2 & 51.8 ± 3.03  & 52.9 ± 3.42 \\
\multicolumn{2}{c}{} & SSF-CNN \cite{cai2021low} & 62.5 ± 3.40 & 69.8 ± 5.12 & 73.3 ± 6.21 & 52.3 ± 3.50  & 53.4 ± 4.16 \\
\multicolumn{2}{c}{} & MBSSFCC \cite{jiang2022detecting} & 66.9 ± 5.00 & 75.6 ± 6.55 & 78.7 ± 6.75 & 52.5 ± 4.35  & 53.9 ± 5.80 \\
\multicolumn{2}{c}{} & EEGraph \cite{cai2023brain} & 72.5 ± 7.41 & 78.7 ± 6.47 & 79.4 ± 7.16 & \multicolumn{1}{c}{-}  & \multicolumn{1}{c}{-} \\
\multicolumn{2}{c}{} & DGSD \cite{fan2024dgsd} & \multicolumn{1}{c}{-} & 79.6 ± 6.76 & 82.4 ± 6.86 & 55.2 ± 4.07  & \multicolumn{1}{c}{-} \\
\multicolumn{2}{c}{} & DBPNet \cite{ni2024dbpnet} & 74.0 ± 5.20 & 79.8 ± 6.91 & 80.2 ± 6.79 & 55.5 ± 6.33  & 55.8 ± 6.11 \\
\multicolumn{2}{c}{} & DARNet \cite{yan2024darnet} & 74.6 ± 6.09 & 80.1 ± 6.85 & 81.2 ± 6.34 & 55.6 ± 4.13  & 55.6 ± 4.04 \\
\multicolumn{2}{c}{} & \textbf{ListenNet (ours)} & \textbf{79.4 ± 7.00} & \textbf{86.2 ± 5.55} & \textbf{86.6 ± 4.82} & \textbf{56.8 ± 7.32}  & \textbf{57.2 ± 5.83} \\

\midrule
\multirow{10}{*}{AVED}  

& \multirow{5}{*}{audio-only} & SSF-CNN \cite{cai2021low}                      & 53.4 ± 1.47                    & 58.4 ± 3.79                  & 58.9 ± 5.35 & 51.7 ± 0.85  & 52.5 ± 1.55 \\
&                                                                     & MBSSFCC \cite{jiang2022detecting}              & 55.9 ± 1.80                    & 70.2 ± 4.10                  & 74.2 ± 7.24  & 52.2 ± 1.52  & 52.7 ± 1.87 \\
&                                                                     & DBPNet \cite{ni2024dbpnet}              & 53.6 ± 2.65                    & 58.9 ± 3.65                  & 62.8 ± 5.93 & 52.1 ± 1.19  & 53.3 ± 1.88 \\
&                                                                     & DARNet \cite{yan2024darnet}              & 49.7 ± 1.05                    & \textbf{80.2 ± 14.67}                  & \textbf{83.6 ± 12.10} & 51.3 ± 0.21  & 52.1 ± 1.54 \\
&                                                                     & \textbf{ListenNet (ours)}                                & \textbf{57.7 ± 1.71}            & 74.6 ± 3.36          & 77.1 ± 5.31         & \textbf{52.8 ± 1.30}   & \textbf{53.8 ± 1.98} \\ \cmidrule{2-8} 
& \multirow{5}{*}{audio-visual}                                        & SSF-CNN \cite{cai2021low}                & 54.5 ± 1.79                    & 59.2 ± 5.44                  & 63.1 ± 6.55            & 52.4 ± 2.29  & 53.8 ± 2.27 \\
&                                                                     & MBSSFCC \cite{jiang2022detecting}         & 57.5 ± 2.75                    & 69.6 ± 5.57                  & 75.5 ± 4.34            & 52.8 ± 1.57  & 54.1 ± 1.86 \\
&                                                                     & DBPNet \cite{ni2024dbpnet}              & 56.1 ± 2.68                    & 61.5 ± 4.33                  & 64.1 ± 6.09   & 53.3 ± 2.39  & 54.0 ± 1.61               \\
&                                                                     & DARNet \cite{yan2024darnet}              & 50.3 ± 0.77                   & \textbf{83.6 ± 12.10}                 & \textbf{88.7 ± 13.15}    & 51.4 ± 0.32    & 52.6 ± 0.29          \\
&                                                                     & \textbf{ListenNet (ours)}                                & \textbf{57.9 ± 2.16}            & 74.9 ± 4.63         & 76.5 ± 5.07        & \textbf{53.7 ± 1.60}                 & \textbf{54.1 ± 1.83}   \\ \bottomrule[1pt]
\end{tabular}
}
\caption{Comparison of accuracy (\%) on KUL, DTU and AVED datasets. The subject-dependent setup is conducted with three decision windows (0.1-second, 1-second, 2-second), with the results for AVED being reproduced, and the remaining results replicated from the corresponding papers. The subject-independent setup is conducted with two decision windows (1-second, 2-second), with DGSD from the original paper and others reproduced. Best results are highlighted in bold.}
\label{tab:2}
\end{table*}
\setlength{\textfloatsep}{15pt} 

In this work, we maintain the same subject-dependent setup as most existing models and evaluate our model in a more challenging subject-independent setup to better align with real-world applications, as detailed in Table \ref{tab:2}. 

\subsubsection{Performance of Subject-Dependent}
The comparison of subject dependence AAD performance between the ListenNet model and other baselines on the KUL, DTU and AVED datasets is presented in Tables \ref{tab:2}. Our method significantly outperforms the current SOTA on both the KUL and DTU datasets. Specifically, on the KUL dataset, ListenNet demonstrates higher accuracies by 3.3\%, 2.1\%, and 1.8\% for the 0.1-second, 1-second, and 2-second decision windows, respectively. Similarly, on the DTU dataset, it achieves improvements of 4.8\%, 6.1\%, and 5.4\% in the same decision windows. 
On the AVED dataset, ListenNet performs slightly worse than DARNet in the 1-second and 2-second decision windows, but still achieves optimal performance in the very short 0.1-second window. One possible explanation is that DARNet's transformer attention outperforming by capturing long-range cross-modal dependencies in the AVED dataset. 

We observe that ListenNet's decoding accuracy increases with the enlargement of decision windows, due to longer decision windows providing more information. The proposed ListenNet exhibits satisfactory performance at a temporal resolution of 1-second, which is approximately close to the time lag necessary for humans to switch attention. Moreover, our advantages are further enhanced under the highly challenging short 0.1-second decision window length, thereby contributing to the subsequent realization of real-time decoding of auditory attention.

\subsubsection{Performance of Subject-Independent}
Apart from excellent results in the subject-dependent setup, the proposed ListenNet also demonstrates comprehensive leading classification performance in the more challenging subject-independent setup across three datasets for the commonly used two detection window sizes. ListenNet benefits from better results by more comprehensively and effectively integrating dynamic temporal patterns and spatio-temporal dependencies, enabling the model to flexibly utilize subject-invariant representations. The results further confirm this capability. Especially on the KUL dataset, ListenNet achieves notable performance, demonstrating accuracy increases of 8.2\% and 7.7\% over the current SOTA model for the 1-second and 2-second decision windows, respectively. Furthermore, ListenNet outperforms baselines for DTU and AVED as well.

Compared to the widely-used KUL dataset, the DTU and AVED datasets pose a more challenging AAD task. Specifically, DTU presents speech at a narrower angle, and its recording environment includes reverberation and background noise, whereas AVED introduces complex multi-modal stimulus materials. The results show that ListenNet outperforms the baseline methods across diverse datasets, with lower variability in its results, further highlighting the stability and reliability of our approach across different decision windows. It learns the common pattern of feature distribution from subjects, thereby more effectively simulating real-world scenarios. These results highlight the robustness and generalization capabilities of the proposed model, emphasizing its potential superiority in EEG-based applications.

\subsection{Ablation Analysis}
\begin{table}[tb]
\centering
\resizebox{0.5\textwidth}{!}{%
\begin{tabular}{@{}cccccc@{}}
\toprule[1pt]
\textbf{Dataset} & \textbf{Model} & \textbf{Subject-Dependent} & \textbf{Subject-Independent} \\ \midrule
\multirow{5}{*}{KUL}             
                                 & w/o STDE-T                               & 91.1 ± 6.05                                    & 62.6 ± 11.10                                    \\
                                  & w/o STDE-S                               & 94.6 ± 6.18                                    & 76.0 ± 15.05                                    \\
                                  & w/o MSTE                               & 96.7 ± 3.46                                    & 77.8 ± 13.39                           \\
                                  & w/o CNA                                & 96.3 ± 2.76                                    & 77.7 ± 14.74                                    \\ 
                                  & \textbf{ListenNet}               & \textbf{96.9 ± 3.01}                          & \textbf{78.1 ± 13.50} \\ \midrule
                                                           
\multirow{5}{*}{DTU}             & w/o STDE-T                               & 72.5 ± 5.53                                    & 52.3 ± 2.01                                    \\
                                  & w/o STDE-S                              & 84.3 ± 5.89                                   & 54.3 ± 8.36                                    \\
                                  & w/o MSTE                               & 84.9 ± 6.59                                    & 56.7 ± 7.91                          \\
                                  & w/o CNA                                & 85.8 ± 5.75                                    & 56.5 ± 5.83                                   \\
                                  & \textbf{ListenNet}               & \textbf{86.2 ± 5.55}                          & \textbf{56.8 ± 7.32}                           \\ \midrule
\multirow{5}{*}{\makecell{AVED \\ (audio-only)}}          & w/o STDE-T                               & 64.2 ± 6.62                                    & 51.1 ± 1.43                                   \\
                                  & w/o STDE-S                               & 66.2 ± 4.50                                    & 52.6 ± 1.71                                    \\
                                  & w/o MSTE                               & 71.8 ± 3.00                                    & 52.5 ± 1.48                                 \\
                                  & w/o CNA                                & 74.3 ± 3.36                          & 52.5 ± 1.32                                  \\
                                  & \textbf{ListenNet}               &  \textbf{74.6 ± 3.36}                                     & \textbf{52.8 ± 1.30}                           \\ \midrule
\multirow{5}{*}{\makecell{AVED \\ (audio-visual)}} & w/o STDE-T                               & 64.9 ± 5.30                                    & 53.3 ± 3.03                                    \\
                                  & w/o STDE-S                               & 66.2 ± 5.27                                    & 53.2 ± 2.14                                   \\
                                  & w/o MSTE                                & 72.8 ± 3.40                                   & 53.6 ± 1.80                                    \\
                                  & w/o CNA                                & 74.6 ± 3.08                          & 53.2 ± 2.53                                    \\
                                  & \textbf{ListenNet}               & \textbf{74.9 ± 4.63}                                    & \textbf{53.7 ± 1.60}                           \\ \bottomrule[1pt]
\end{tabular}%
}
\caption{Ablation study on all three datasets. The subject-dependent and subject-independent setups are conducted with 1-second decision windows, and “w/o” means without.}
\label{tab:3}
\end{table}

Ablation studies are conducted on three datasets using a 1-second window setting, which most closely aligns with human attention switching \cite{jiang2022detecting,fan2025seeing}. ListenNet constructs robust spatio-temporal representations. This enables the model to capture the full spatiotemporal information in EEG signals, thereby improving the interpretation of brain activity. Table \ref{tab:3} presents a comparison between the full ListenNet model and these four variants across the three datasets.

STDE-T and STDE-S are each removed to disrupt the integrity of STDE, thereby assessing the critical role of these components in the model's performance. Removing the STDE-T module for spatio-temporal dependency encoding has the most significant impact on the model's performance. The effectiveness of STDE-T can be attributed to the fact that EEG signals, as high temporal-resolution time series, exhibit strong temporal dependencies. Prioritizing the modeling of temporal continuity allows for the extraction of more effective and accurate spatio-temporal feature embeddings. Removing the STDE-S module results in accuracy decline, as full-channel spatial convolution captures inter-channel dependencies and establishes a robust spatio-temporal feature framework. 
 
The removal of the MSTE module leads to the loss of multi-scale temporal information and the disruption of potential dependencies between temporal segments, increasing the risk of missing critical temporal features. Finally, removing the CNA module eliminates the model's ability to dynamically assign feature weights and enhance spatio-temporal representations, thereby weakening the extraction and integration of multi-level spatio-temporal features and further reducing accuracy.

\subsection{Computational Cost}

\begin{table}[tb]
\centering
\resizebox{0.9\columnwidth}{!}{
\begin{tabular}{ccc}  
\toprule[1pt]
\textbf{Model} & \textbf{Params (M)} & \textbf{MACs (M)} \\ \midrule
MBSSFCC \cite{jiang2022detecting} & 83.91 & 89.15 \\
DBPNet \cite{ni2024dbpnet} & 0.91 & 96.55\\
DARNet \cite{yan2024darnet} & 0.08 & 16.36 \\
\textbf{ListenNet (ours)} & \textbf{0.01} & \textbf{12.16} \\ 
\bottomrule[1pt]
\end{tabular}
}
\caption{ The training parameter counts (Params) and multiply-accumulates (MACs) comparison on the KUL dataset.}
\label{tab:5}
\end{table}

\begin{figure}[tb]
\centering
\includegraphics[width=1\columnwidth]{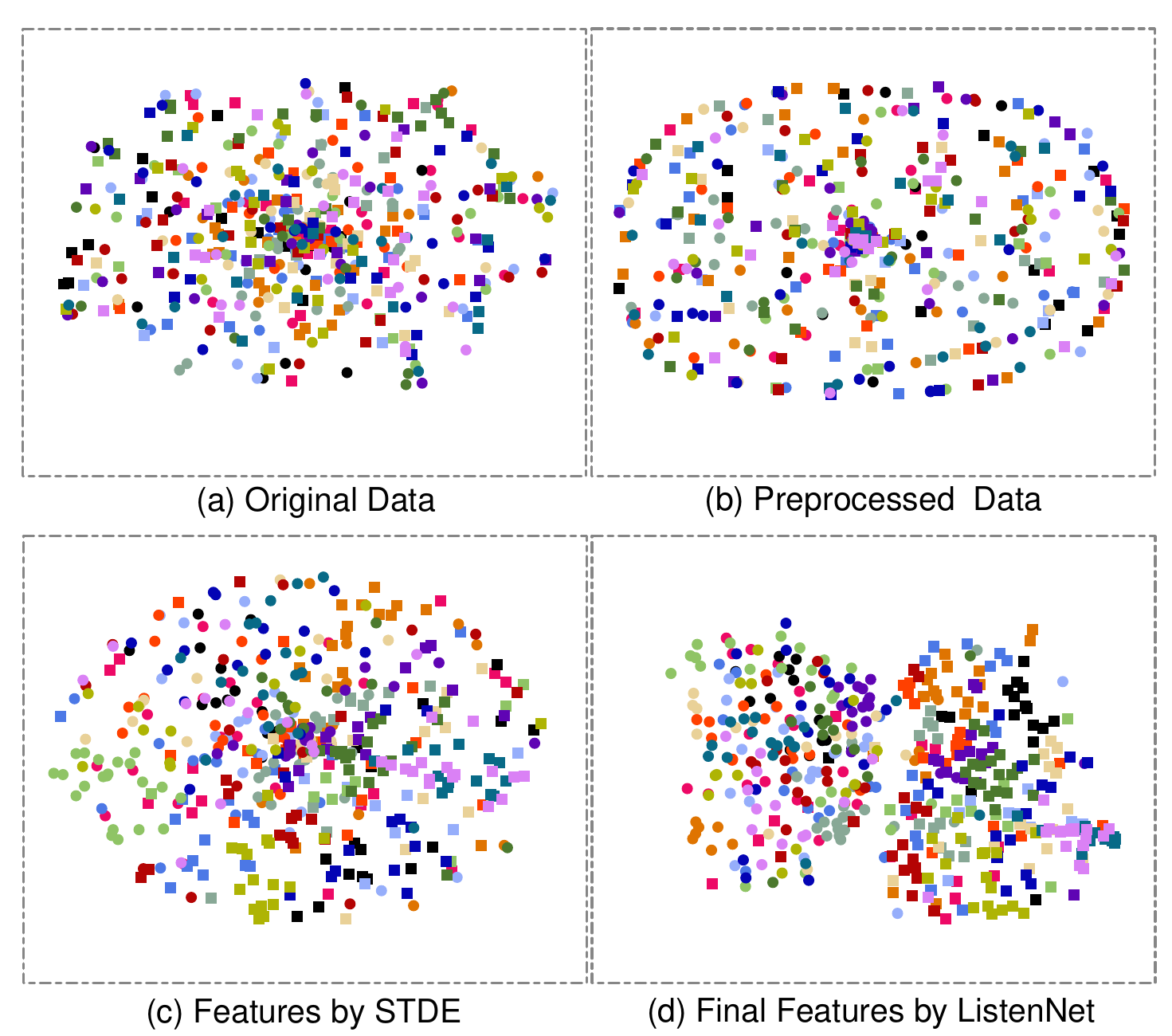}
\caption{The t-SNE visualization of different types of features on the KUL dataset under the subject-independent condition. Different colors represent different subjects. Circles and squares denote attention to the left or right speaker, respectively.}
\label{fig:4}
\end{figure}


Table \ref{tab:5} compares the parameter counts and MACs of ListenNet with those of MBSSFCC, DBPNet, and DARNet on the KUL dataset. With only 0.01 M trainable parameters, ListenNet achieves remarkable parameter efficiency, requiring approximately 8390 times fewer parameters than MBSSFCC, 90 times fewer parameters than DBPNet, 7 times fewer parameters than DARNet. Additionally, ListenNet's computational demand is also markedly reduced, with its MACs only 12.16 M, approximately 86\% lower than MBSSFCC, 87\% lower than DBPNet and 26\% lower than DARNet. These substantial reductions in both parameter count and computational complexity highlight ListenNet's enhanced efficiency, making it especially suitable for deployment on devices with limited computational resources.

\subsection{Visualization Analysis}
To assess the effect of extracting subject-invariant features, we randomly select 30 samples from each subject in the KUL dataset and visualize them using t-SNE \cite{van2008visualizing}. The resulting plots are shown in Figure \ref{fig:4}. Different colors represent subjects, with circles and squares indicating attention to the left or right speaker, respectively. In Figure \ref{fig:4}~(a), the raw features are scattered with significant overlap between subjects and labels, lacking clear structure and separability. In Figure \ref{fig:4}~(b), preprocessing improves feature quality to some extent, but notable overlap and insufficient separability still remain. In Figure \ref{fig:4}~(c), features extracted using STDE form clearer attention-related subgroups. By capturing spatio-temporal cross dependencies, the STDE module learns dynamic patterns and enhances feature separability, though some class boundaries remain indistinct. In Figure \ref{fig:4}~(d), features extracted by ListenNet exhibit more distinct clustering for attention labels across subjects, and the distributions become more organized. This demonstrates that ListenNet learns subject-invariant features while maintaining clear boundaries between attention categories.


\section{Conclusion}
This paper introduces ListenNet, a lightweight, highly accurate, and generalizable network for AAD. By combining spatio-temporal convolution operations across time steps and all channels, it effectively utilizes spatial information embedded in temporal EEG signals. Additionally, it captures temporal patterns at multiple scales, previously overlooked, by using multi-scale dilated convolutions. It integrates hierarchical spatio-temporal features through cross-nested attention mechanisms. Subject-dependent and subject-independent experiments are conducted on three AAD datasets. Experimental results show that our ListenNet exhibits competitive accuracy, especially in the very short 0.1-second decision window and across subjects. Furthermore, the compact size of our model and the reduced computational costs open new possibilities for deployment on low-power devices. For future work, we intend to extend ListenNet to streaming architectures, integrating incremental learning for real-time adaptation to AAD scenarios.

\section*{Acknowledgements}
This work is supported by the {STI 2030—Major Projects (No.2021ZD0201500)}, the National Natural Science Foundation of China (NSFC) (No.62201002, 6247077204), Excellent Youth Foundation of Anhui Scientific Committee (No. 2408085Y034), Distinguished Youth Foundation of Anhui Scientific Committee (No.2208085J05), Special Fund for Key Program of Science and Technology of Anhui Province (No.202203a07020008), Cloud Ginger XR-1.

\bibliographystyle{named}
\bibliography{ijcai25}

\end{document}